\documentclass[conference]{IEEEtran}
\usepackage{balance}  

\usepackage{supertabular}
\usepackage{makecell, multirow, tabularx}
\usepackage[noprefix, nomentbl]{nomencl}
\usepackage[draft]{hyperref}
\usepackage{cite}
\usepackage{amsmath,amssymb,amsfonts}
\usepackage{algorithmic}

\usepackage{graphicx}
\usepackage{textcomp}
\usepackage{xcolor}
\usepackage{amsmath}
\usepackage{algorithm,algorithmic}
\usepackage{cellspace}    
\usepackage{makecell, multirow, tabularx}
\usepackage{subcaption,graphicx}
\usepackage{pbox}
\usepackage{amssymb}
\usepackage{pifont}
\usepackage{verbatim}
\usepackage{algorithm}
\usepackage{algorithmic}
\usepackage{tabularx}
\usepackage{graphicx}
\usepackage{multirow}
\usepackage{tcolorbox}
\usepackage{xcolor}
\usepackage{listings}

\definecolor{terminalbackground}{RGB}{48,10,36}
\definecolor{terminalgreen}{RGB}{0,255,0}
\definecolor{terminalred}{RGB}{255,0,0}

\lstdefinestyle{terminal}{
    basicstyle=\small\ttfamily\color{white},
    backgroundcolor=\color{terminalbackground},
    breaklines=true,
    captionpos=b,
    keepspaces=true,
    numbers=none,
    numbersep=5pt,
    showspaces=false,
    showstringspaces=false,
    showtabs=false,
    tabsize=2,
    moredelim=[is][\color{terminalgreen}]{<g>}{</g>},
    moredelim=[is][\color{terminalred}]{<r>}{</r>},
}

\usepackage[skip=1pt]{caption}
\usepackage{enumitem}

\usepackage[british]{babel}
\def\BibTeX{{\rm B\kern-.05em{\sc i\kern-.025em b}\kern-.08em
    T\kern-.1667em\lower.7ex\hbox{E}\kern-.125emX}}
    \usepackage{mathtools}

\usepackage[top=0.69in,bottom=1.53in,right=0.6in,left=0.6in]{geometry}

\usepackage{lipsum}

\usepackage[font=small,skip=1pt]{caption}

\begin{document}



\title{LPUF-AuthNet: A Lightweight PUF-Based IoT Authentication via Tandem Neural Networks and Split Learning\\



{\footnotesize \textsuperscript{}}

}

\DeclareRobustCommand{\IEEEauthorrefmark}[1]{\smash{\textsuperscript{\footnotesize #1}}}




\author{
    \IEEEauthorblockN{Brahim Mefgouda$^{1,2}$, Raviha Khan$^{2}$, Omar Alhussein$^{1,3}$, Hani Saleh$^{2}$, Hossien B. Eldeeb$^{1,2}$, \\Anshul Pandey$^{4}$, and Sami Muhaidat$^{1,2,5}$}
    \IEEEauthorblockA{$^{1}$KU 6G Research Center, Khalifa University, Abu Dhabi, UAE\\
                      $^{2}$Department of Communication and Information Systems, Khalifa University, Abu Dhabi, UAE\\
                      $^{3}$Department of Computer Science, Khalifa University, Abu Dhabi, UAE\\
                      $^{4}$Secure Systems Research Center, Technology Innovation Institute, Abu Dhabi, UAE\\       
                      $^{5}$Department of Systems and Computer Engineering, Carleton University, Ottawa, Canada\\
                      Emails: \{brahim.mefgouda, 100060567, omar.alhussein, hani.saleh\}@ku.ac.ae, \\hossieneldeeb@ieee.org, anshul.pandey@tii.ae, and muhaidat@ieee.org}
}

\maketitle
{\color{black}
\begin{abstract}
By 2025, the internet of things (IoT) is projected to connect over 75 billion devices globally, fundamentally altering how we interact with our environments in both urban and rural settings. However, IoT device security remains challenging, particularly in the authentication process. Traditional cryptographic methods often struggle with the constraints of IoT devices, such as limited computational power and storage. This paper considers physical unclonable functions (PUFs) as robust security solutions, utilizing their inherent physical uniqueness to authenticate devices securely. However, traditional PUF systems are vulnerable to machine learning (ML) attacks and burdened by large datasets. Our proposed solution introduces a lightweight PUF mechanism, called LPUF-AuthNet, combining tandem neural networks (TNN) with a split learning (SL) paradigm. The proposed approach provides scalability, supports mutual authentication, and enhances security by resisting various types of attacks, paving the way for secure integration into future 6G technologies. 
\end{abstract}}
\thispagestyle{empty}
\pagestyle{empty}

\section{Introduction}
The internet of things (IoT) has become integral to the modern world, offering seamless connectivity, automation, and intelligent decision-making to make routine and time-consuming tasks more convenient. The widespread adoption of IoT technology is evident from the staggering number of connected devices worldwide. 
Currently, around 30 billion IoT devices are in use, with projections indicating this number will reach 75 billion by 2025 \cite{jamshed2022challenges}.  However, the heterogeneity of interconnected devices in IoT ecosystems poses significant security challenges, necessitating robust protection mechanisms \cite{hassija2019survey}. The security architecture must account for the diverse capabilities of the intelligent infrastructure (nodes) to ensure secure communications. A cornerstone of IoT security is the authentication process, which verifies the identities of devices and users to prevent unauthorized access. Since IoT devices often handle and transmit sensitive data, they are highly vulnerable to cyberattacks that threaten user privacy and safety. Traditional authentication and key exchange methods, including those based on symmetric and asymmetric cryptography, have been deployed to counter various threats, such as replay, man-in-the-middle (MITM), and cloning attacks. However, integrating these cryptographic schemes into IoT systems presents considerable challenges due to the substantial computational, storage, and communication overhead they introduce \cite{zheng2022puf}.


Alternatively, physical unclonable functions (PUFs), based on inherent randomness in manufacturing and material properties, are unique hardware-based security primitives that generate distinct challenge-response pairs (CRPs) \cite{pappu2002physical,saleh4838393encoder}. Each PUF produces a unique digital fingerprint due to the variability in responses for the same challenge, making them highly effective for authentication. However, PUF-based schemes often require the storage of extensive CRP datasets, which poses significant challenges for IoT devices with limited storage and memory resources \cite{mall2022puf}. Additionally, integrating PUF security into IoT systems is constrained by hardware limitations, cost considerations, and the risk of side-channel attacks. Studies have demonstrated the potential for cloning PUFs using side-channel analysis and machine learning (ML) techniques, raising concerns about their long-term security resilience \cite{lohachab2020comprehensive}.

To address such implementations with PUFs in the context of IoT, Nimmy \textit{et al.} propose an authentication framework that integrates geometric threshold secret-sharing with PUFs \cite{nimmy2023novel}. This approach eliminates the need to store the CRP dataset in the verifier node by dividing each CRP into shares and only storing these shares and a hash of the response. Moreover, Chatterjee \textit{et al.} develop an authentication and key exchange protocol that combines the ideas of identity-based encryption and keyed hash functions to eliminate the need for explicit CRP storage in the verifier database \cite{chatterjee2018building}. Also, Zhang \textit{et al.} introduce a PUF-as-a-service framework using shamir secret sharing (SSS) and blockchain to protect CRPs \cite{zhang2024building}. This approach avoids storing CRPs in the verifier by distributing CRP shares across multiple providers. Furthermore, the authors in \cite{harishma2020safe} also avoid storing CRPs in datasets by using identity-based encryption, with the CRPs stored in a cloud server to assist different verifiers in authenticating smart meters.

However, the aforementioned solutions \cite{nimmy2023novel}-\cite{harishma2020safe} have several limitations. For instance, using complex secret-sharing schemes or a verifier helper brings additional computational complexity, which can significantly burden the processing part on the already resource-constrained IoT devices. Additionally, the high number of messages exchanged between the verifier and the node undergoing authentication can lead to increased communication costs and latency. Moreover, certain protocols continue to necessitate the storage of a subset of CRPs, which may impede scalability in dynamic IoT ecosystems characterized by high device density. Furthermore, existing works do not address threats, such as forward secrecy (FS) \cite{zhang2024building}, mutual authentication (MA) \cite{harishma2020safe},  MITM attacks \cite{nimmy2023novel}, and ML-based attacks \cite{chatterjee2018building}. Accordingly, this paper proposes a novel lightweight PUF authentication scheme termed LPUF-AuthNet, which comprises two ML models: deep neural networks (DNN) and tandem neural networks (TNN) trained using split learning (SL) paradigm \cite{vepakomma2018split,alhussein2023dynamic}. The TNN is a neural network (NN) architecture characterized by the sequential connection of an inverse design network to a forward modeling network \cite{liu2018training}. The contribution of this paper can be summarized as follows:

{\color{black}
\begin{itemize}
    \item We develop novel machine-learning models that emulate the behaviour of hardware PUFs in generating CRPs and accurately predicting responses to corresponding challenges. This approach eliminates the dependency on physical PUFs and removes the need to store large CRP datasets in the verifier node.
    \item  We present a method that leverages the developed ML models to encode challenges into a compact latent challenge (LC) form and decode latent responses (LRs) from legitimate nodes. TNN model is composed of two collaborative blocks: TNN$_1$ and TNN$_2$, where each block is composed of an encoder and a decoder. TNN$_1$ operates on the verifier, which is responsible for validating the legitimacy of the responses.  In parallel, TNN$_2$ is implemented on the legitimate nodes, where it verifies the authenticity of incoming challenges. This dual-verification mechanism ensures bidirectional security, thereby significantly enhancing the integrity and robustness of the authentication protocol.
    \item Analytically, we have shown the effectiveness of the proposed LPUF-AuthNet authentication framework against several security attacks. Furthermore, we showed that the proposed method could very accurately differentiate between real and fake latent space challenges. Also, a comparison study from the existing literature is presented, showing the approach's effectiveness in terms of the communication overhead. We developed a real-time proof-of-concept for the said approach and validated the findings. 
 \end{itemize}
 }
    
    

\section{The Proposed LPUF-AuthNet Protocol}
\subsection{System Overview}
\begin{figure}[!htb]
\centering
\subfloat[]{\includegraphics[width=1\linewidth]{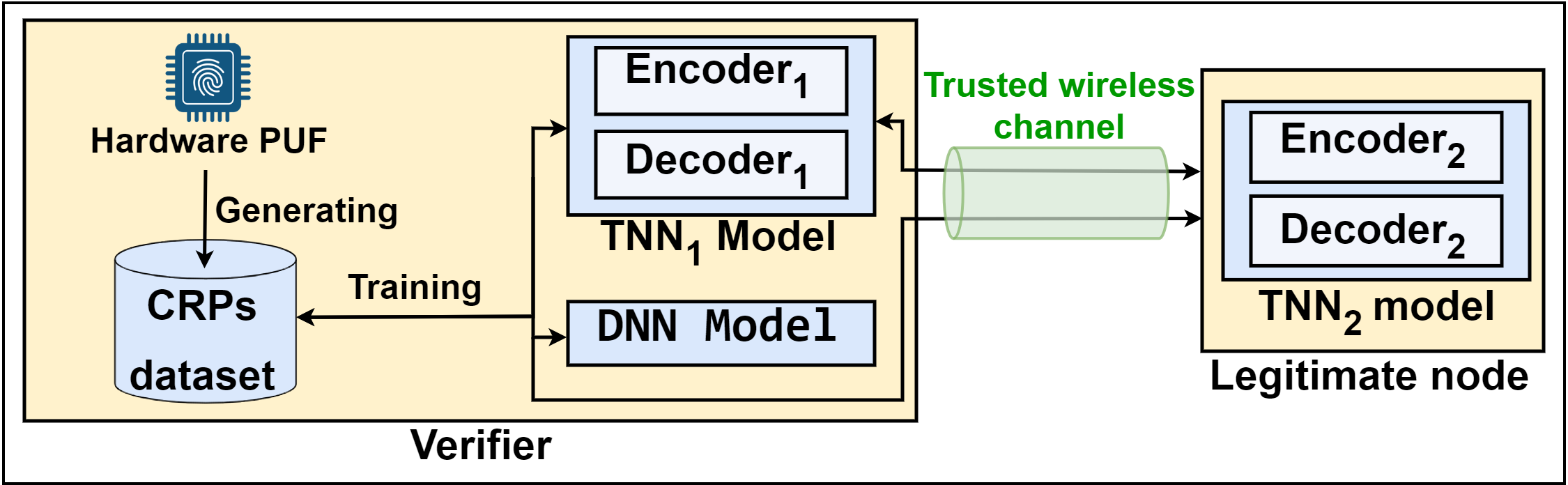}}\\
\subfloat[]{\includegraphics[width=1\linewidth]{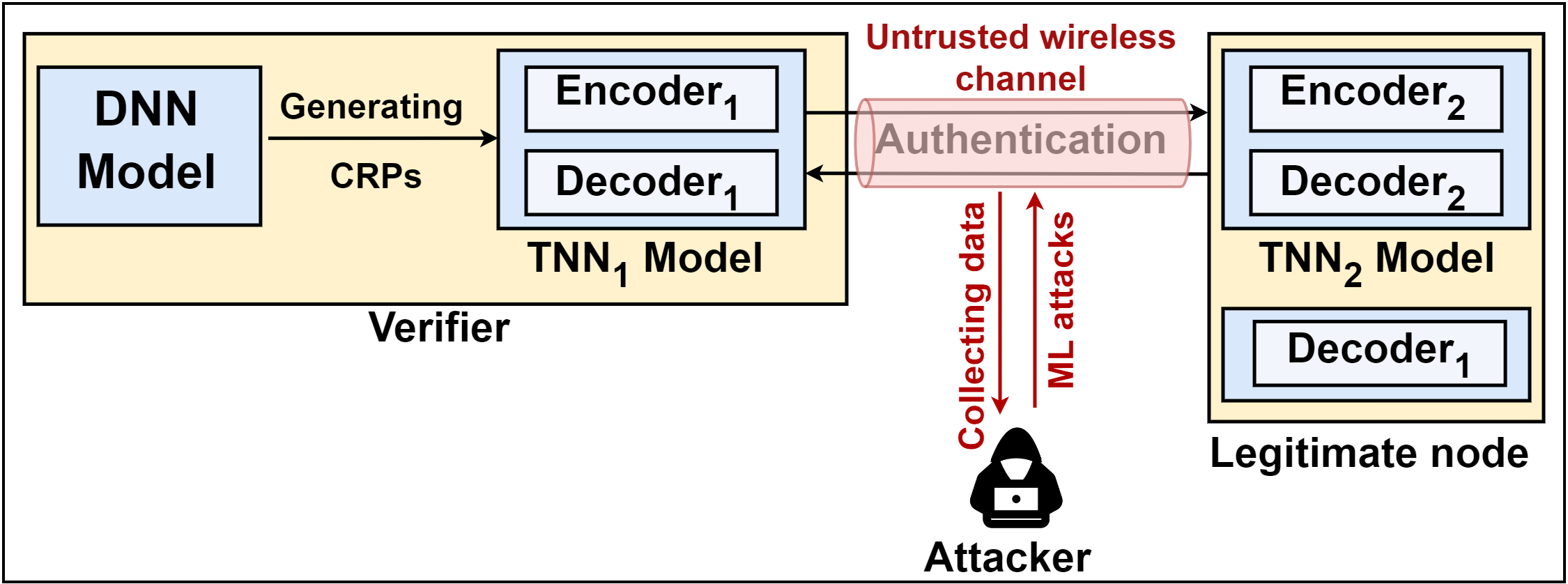}}
    \caption{The proposed system model: (a) Enrollment phase, (b) Authentication phase.}
  \label{s33}
\end{figure}

The system model comprises three entities: legitimate nodes (IoT devices), an authenticator (or verifier), and an attacker, as shown in Fig. \ref{s33}. Each IoT device seeks to authenticate itself through the authenticator, which can be either an IoT device or a standard computing device.
The adversarial entity ($Att$) can intercept, replay, eavesdrop on, and alter messages transmitted through an untrusted wireless channel during the authentication phase. The primary objective of $Att$ is to authenticate itself to the verifier without possessing legitimate credentials. We assume that  $Att$ can only listen to the untrusted wireless channel, and cannot hack the legitimate node or the verifier. Furthermore, we assume that the authenticator node is trustworthy. \textcolor{black}{The provisioning and enrollment phases are assumed to be secure, as non-lightweight techniques (such as public/private-key cryptography can be used)}. 

Here,  $Att$ may attempt various attacks, such as replaying intercepted messages to deceive the verifier, tampering with message content to inject false information or cause disruption, or eavesdropping to gather information to impersonate a legitimate node. These scenarios highlight the importance of implementing robust security measures to protect the authentication process in untrusted environments.







%
The LPUF-AuthNet architecture integrates two NN models to create a robust authentication framework: a DNN and a TNN. The DNN, located in the verifier device, generates new CRPs, eliminating the need for hardware PUFs or pre-stored CRP datasets. The verifier's TNN, TNN$_1$, encodes the challenges into compact LCs while decoding the LRs into responses. Conversely, the TNN residing in legitimate nodes, TNN$_2$, processes incoming LCs from the verifier and predicts corresponding LRs. This system facilitates efficient challenge-response operations and enables MA: the verifier's TNN$_1$ authenticates legitimate nodes, whereas the legitimate node's TNN$_2$ enables nodes to verify the verifier's authenticity. By leveraging the strengths of different NN architectures, LPUF-AuthNet achieves a dynamic, secure, and efficient device authentication process.

The proposed protocol consists of two main stages, as illustrated in Fig. \ref{s33}. \textit{\textbf{a}}) Enrollment: This stage is dedicated to training the ML models that compose the LPUF-AuthNet protocol on CRPs collected from the hardware PUF. \textit{\textbf{b}})  Authentication: This stage involves initiating and conducting authentication between the verifier and legitimate devices.

\subsection{Enrollment Phase: Defining the LPUF-AuthNet architecture}
\subsubsection{\textbf{DNN Model for Generating CRPs}} 
This model is designed to accurately reproduce the original CRPs collected from a hardware PUF while significantly reducing storage requirements. Unlike conventional approaches necessitating substantial memory for CRP datasets, our architecture enables practical implementation in IoT environments without compromising the security inherent to PUF-based systems.

The DNN model takes a binary vector representing the index of a specific CRP from the dataset as input and outputs the corresponding CRP.  {\color{black}Based on extensive experimental evaluations}, the architecture of this model comprises an input layer, five hidden layers with progressively increasing neuronal density (64, 128, 256, 512, and 1024), and an output layer. The input layer size is $\lceil \log_2(\textmd{$n$}+1) \rceil$ ($n$ denotes the size of the CRP dataset),  representing the maximum number of bits needed for the CRP indices, while the output layer size combines the challenge and response bits.

\subsubsection{\textbf{TNN Models for Authentication}}

\begin{figure}[!htbp]
\begin{center}
\includegraphics[width= 0.8\linewidth]{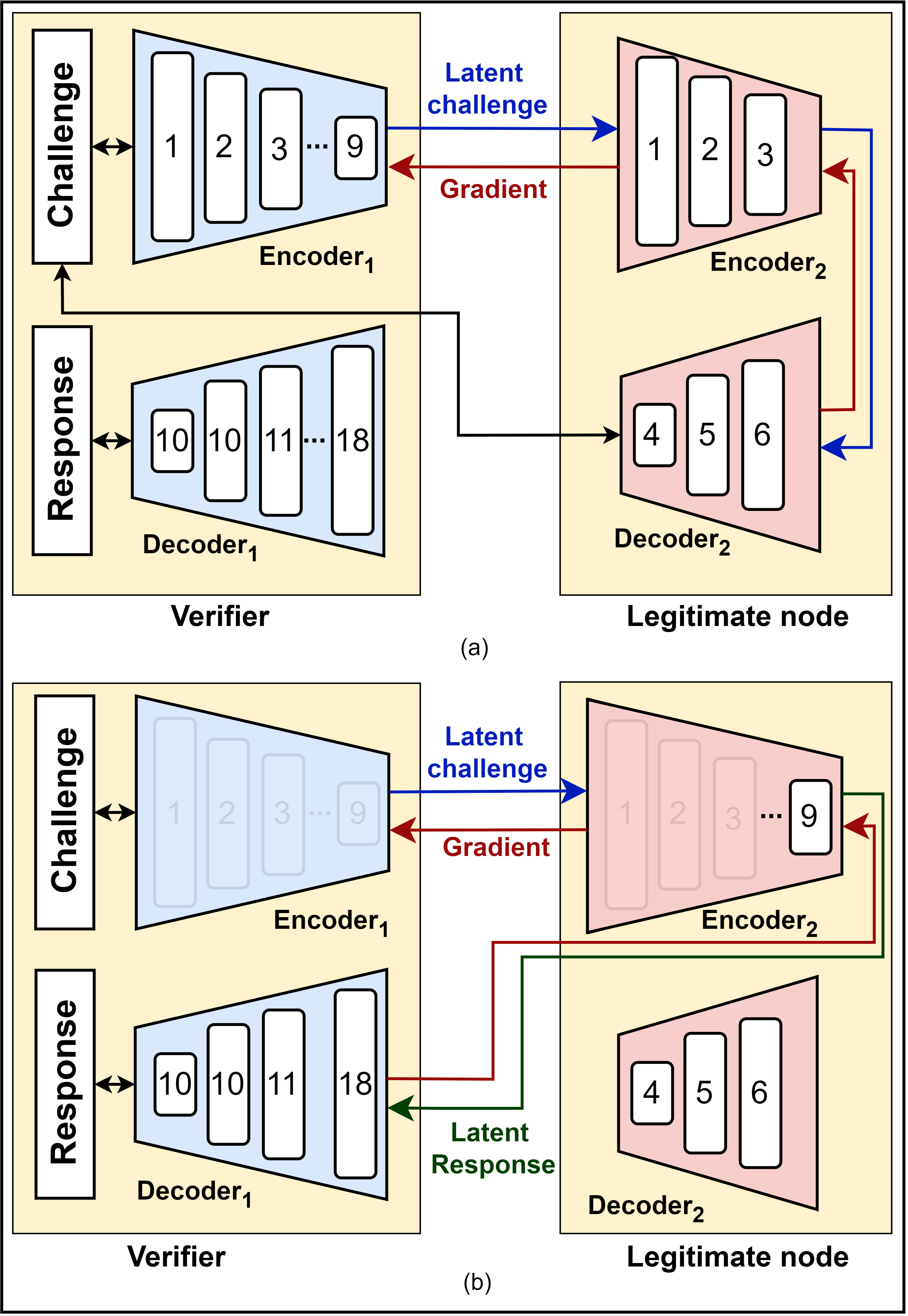}
\caption{Training the TNN models.}
\label{VHOPicture2}       
\end {center}
\end{figure}

%
%
\paragraph{ \textbf{TNN$_1$ Model}}
This model, depicted in Fig. \ref{VHOPicture2}, is fragmented into two components: $Encoder_1$ and $Decoder_1$, each fulfilling a crucial role in the system. $Encoder_1$ compresses the challenge generated by the DNN model into a 4-byte LC for efficient transmission. Conversely, $Decoder_1$ at the verifier is responsible for reconstructing the corresponding response from the received latent representation. This process of compression and reconstruction forms the core of the TNN$_1$'s functionality, enabling efficient communication while maintaining the integrity of the challenge-response mechanism.

The architecture of $Encoder_1$ consists of an input layer, seven hidden layers, and an output layer. The input layer's dimensionality matches the input challenge's length, with each bit represented by a single neuron. The output layer comprises four neurons, corresponding to the 4-byte LC. This structure facilitates the efficient compression of the challenge into a compact latent representation while preserving essential information for subsequent reconstruction.  Mirroring this structure, $Decoder_1$ consists of an input layer, seven hidden layers, and an output layer. In contrast, its input layer is designed to accept the 4-byte LR representation, while its output layer reconstructs the full response. The detailed architectural specifications of both $Encoder_1$ and $Decoder_1$ are presented in Table \ref{tab:CANN_model}.

\begin{table}[!htb]
    \centering
    \caption{Architecture of the proposed TNN$_1$ model.}
    \renewcommand{\arraystretch}{1.3} 
    \begin{tabularx}{\columnwidth}{|c|X|X|X|}
        \hline
        & \textbf{Layer} & \textbf{Number of Neurons} & \textbf{Activation Function} \\
        \hline
        \multirow{3}{*}{\parbox[t]{5mm}{\rotatebox[origin=c]{90}{Encoder$_1$}}} & Input Layer & $\lceil \log_2(n+1) \rceil$ & ReLU \\
        \cline{2-4}
        & Hidden Layers & 1024, 512, 256, 128, 64, 32, 16, 8 & ReLU \\
        \cline{2-4}
        & Output Layer & 4 & Linear \\
        \hline
        \multirow{3}{*}{\parbox[t]{5mm}{\rotatebox[origin=c]{90}{Decoder$_1$}}} & Input Layer & 4 & ReLU \\
        \cline{2-4}
        & Hidden Layers & 8, 16, 32, 64, 128, 256, 512, 1024 & ReLU \\
        \cline{2-4}
        & Output Layer & $size(Response)$ & Linear \\
        \hline
    \end{tabularx}
    \label{tab:CANN_model}
\end{table}

\paragraph{\textbf{TNN$_2$ Model}}
This model serves dual purposes in the authentication process, comprising two essential components: $Encoder_2$ and $Decoder_2$. The primary objective of TNN$_2$ is to verify the authenticity of the LC sender. Concurrently, its secondary objective is to predict the LR corresponding to the received LC. To achieve the primary objective, the architecture of $Encoder_2$ is bifurcated to accommodate the dual roles of TNN$_2$. Its base configuration consists of an input layer with four neurons corresponding to the 4-byte LC and two hidden layers. The output of this initial structure serves as input for $Decoder_2$, facilitating the prediction of the response for MA purposes.  

To address its second objective, $Encoder_2$ incorporates an extended architecture by adding six additional hidden layers while maintaining the input and the hidden layers. Crucially, the weights of the base architecture are frozen after training, ensuring the integrity of the MA process while allowing for the development of response prediction capabilities.
The architecture of $Decoder_2$ mirrors that of $Encoder_2$, predicting the challenge from the output of $Encoder_2$. Table \ref{tab:AE_model} details the architectural specifications of $Encoder_2$ (both base and extended versions) and $Decoder_2$.

\begin{table}[!htb]
\centering
\caption{Architecture of the proposed TNN$_2$ model.}
\renewcommand{\arraystretch}{1.3}

\begin{tabularx}{\columnwidth}{|c|c|X|X|X|}
\hline
\multicolumn{2}{|c|}{\textbf{Component}} & \textbf{Layer} & \textbf{Number of Neurons} & \textbf{Activation Function} \\
\hline
\multirow{7}{*}{\parbox[t]{5mm}{\rotatebox[origin=c]{90}{\makecell{Encoder$_2$}}}}
& \multirow{3}{*}{\parbox[t]{5mm}{\rotatebox[origin=c]{90}{\makecell{Basic \\ Encoder$_2$}}}}
& Input Layer & 4 & ReLU \\
\cline{3-5}
& & Hidden Layers & 1024, 512 & ReLU \\
\cline{3-5}
& & Output Layer & 256 & Linear \\
\cline{2-5}
& \multirow{3}{*}{\parbox[t]{5mm}{\rotatebox[origin=c]{90}{\makecell{Extended\\Encoder$_2$}}}}
& Input Layer & 4 & ReLU \\
\cline{3-5}
& & Hidden Layers & 64, 32, 16, 8 & ReLU \\
\cline{3-5}
& & Output Layer & 4 & Linear \\
\hline
\multicolumn{2}{|c|}{\multirow{3}{*}{\rotatebox[origin=c]{0}{Decoder$_2$}}}
& Input Layer & 256 & ReLU \\
\cline{3-5}
\multicolumn{2}{|c|}{} & Hidden Layers & 512, 256 & ReLU \\
\cline{3-5}
\multicolumn{2}{|c|}{} & Output Layer & 32 & Linear \\
\hline
\end{tabularx}
\label{tab:AE_model}
\end{table}

{\color{blue}
\paragraph{\textbf{SL for TNN Models} } }
TNN models are trained in two phases using SL, preserving data privacy, as illustrated in Fig. \ref{VHOPicture2}. In the initial phase (a), $Encoder_1$ is trained to generate challenge representations, LC, while the TNN$_2$ learns to differentiate between genuine LCs from $Encoder_1$ and potential forgeries. Consequently, $Encoder_1$, $Encoder_2$, and $Decoder_2$ are trained jointly. Challenges from the dataset are fed to $Encoder_1$, which produces an LC. This LC is then securely transmitted to the legitimate node housing the TNN$_2$, which reconstructs the challenge from the LC. This method pushes model overfitting and retention of CRPs during training, enabling the TNN$_2$ to identify learned challenges. During training, $Decoder_2$ sends gradients back to $Encoder_2$, which then transmits gradients to $Encoder_1$ for updating their weights and those of the TNN$_2$.

In the second phase (b), $Decoder_2$ learns to predict responses from the LR received from $Encoder_2$ hosted in legitimate nodes. To preserve the trained weights of $Encoder_1$ from phase (a), which mapped challenges to LC, all layers of $Encoder_1$ are frozen during this phase. 

We extend the layers of $Encoder_2$ by adding new layers to adapt it for interaction with $Decoder_2$. This extended $Encoder_2$ generates LR corresponding to the LC received from $Encoder_1$. During training, $Encoder_2$ receives LC from $Encoder_1$ and learns to generate LR, which is then transmitted to $Decoder_1$ hosted in the verifier node. The verifier node learns to decompress the LR to generate a response. Gradients are subsequently backpropagated from $Decoder_1$ to $Encoder_2$ and then to $Encoder_1$.

After training the TNN models, the verifier sends a copy of the model's $Decoder_1$ to the legitimate nodes. This helps them verify whether the challenge is received from a trusted verifier.
\begin{figure}[!htb]
\begin{center}
\includegraphics[width= 1\linewidth]{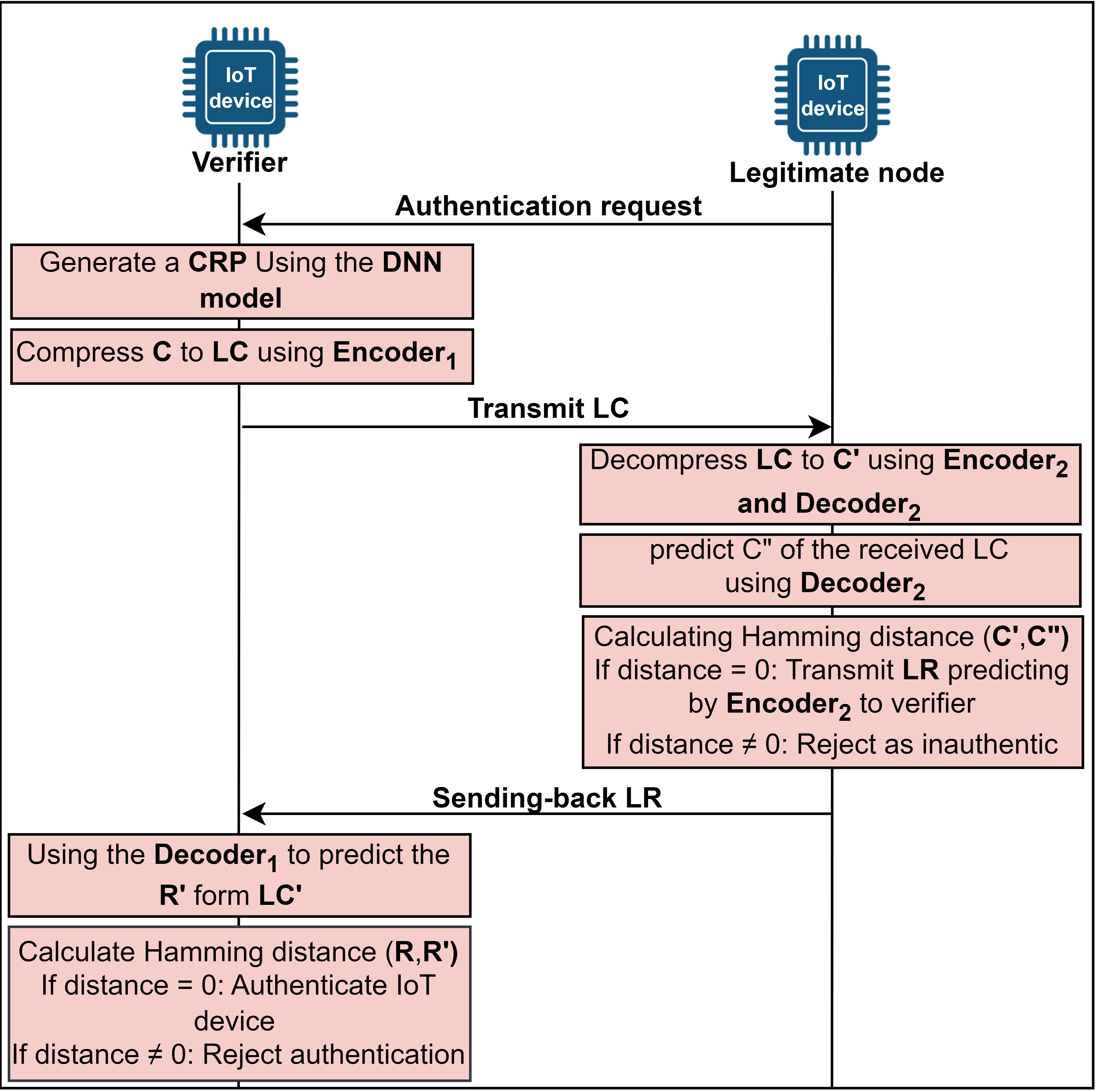}
\caption{LPUF-AuthNet authentication.}
\label{Authentication}       
\end {center}
\end{figure}

{\color{black} The LPUF-AuthNet framework demonstrates scalability in multi-user scenarios, a critical feature for large-scale IoT deployments. In such scenarios, a single hardware PUF is employed to train the DNN, ensuring efficient utilization of resources. Each legitimate IoT node is provisioned with its own instance of TNN$_2$, while the verifier maintains both the DNN and TNN$_1$. This architecture enables efficient and secure authentication for a growing number of devices without compromising performance or security. To integrate a new legitimate IoT node, the system simply provisions it with an instance of TNN$_2$, allowing seamless network expansion. As a result, LPUF-AuthNet provides a robust and flexible solution that adapts to the dynamic nature of large-scale IoT deployments while preserving its core security features and operational efficiency.
}

\subsection{Authentication Phase} 

The authentication phase begins when an IoT node ($N$) sends an authentication request to the verifier ($A$), as illustrated in Fig. \ref{Authentication}. This request indicates that $N$ seeks authentication to establish a secure connection or gain access to specific resources. In response to the authentication request, the verifier uses the DNN to generate a new CRP $(C,R)$ using a previously unused random number between zero and $n-1$. The response $R$ is temporarily stored, and the challenge $C$ is fed into $Encoder_1$ to be represented as a LC, which is then sent to node $N$. A timer $t$ is started by the authenticator $A$, requiring node $N$ to respond within the maximum allowable time $T_{max}$.

After receiving the LC, $N$ performs two parallel operations: \textit{\textbf{a}}) Reconstructing the $C'$ using the TNN$_2$ model, \textit{\textbf{b}}) Generating $C"$ using $Encoder_2$ situated within $N$. Then, Node $N$ computes the hamming distance $H(C', C")$. If $H(C', C") \neq 0$, node $N$ terminates the authentication process due to a potential security attack. Conversely, if $H(C', C") = 0$, node $N$ processes the LC through $Encoder_2$ to generate a LR, which is then sent back to node $A$. Upon receiving the LR,  $A$ decodes it using $Decoder_2$ to retrieve the corresponding  $R'$. The authenticator then calculates the hamming distance between the stored $R$ and $R'$. If the distance is zero, node $N$ is authenticated as legitimate. Otherwise, node $N$ is considered an $Att$ and denied authentication.

\section{Security Analysis}
This section presents a comprehensive security analysis of the proposed protocol as outlined below: 

\paragraph{FS Attacks} 
The DNN model generates a new CRP for each session and ensures that each CRP is used only once. Thus, compromising either the challenge or the response does not compromise previously established sessions. Additionally, each session introduces a unique CRP, ensuring that the protocol maintains FS.

\paragraph{MA Attacks}
Once node $N$ requests authentication, it receives a $LC$. Node $N$ calculates $C'$ and $C''$ using the TNN$_2$ and $Decoder_2$ to verify if the sender of $LC$ is the verifier. Conversely, after node $A$ receives the $LR$ from node $N$, it predicts the response $R'$ and verifies its correctness. Consequently, our model ensures the MA.

\paragraph{MITM Attacks} In this attack, The $Att$ intercepts communications between two parties. Three aspects should be analyzed:
\begin{itemize}
    \item  $Att$ masquerades as $A$: In this case, $N$ can identify that the received $LC$ is not from the authenticator by computing the hamming distance between the received $C'$ and $C"$ generated by the TNN$_2$ model and $Decoder_2$.     
    \item  $Att$ masquerades as $N$: The $Att$ attempts to create an LR to the LC. Upon receiving the LC, $A$ can detect an illegitimate node by employing  $Encoder_2$, $Decoder_2$, and $Decoder_1$, and then calculating the hamming distance. A non-zero distance indicates an unauthorized entity.
    \item  $Att$  captures the response from $N$: By measuring the time to capture and send the $LC$ to $A$. We observe that the authenticator timer $t$ exceeds the normal message transmission time, indicating $Att$ is not legitimate.
\end{itemize}

\paragraph{Replay Attacks (RA)} 
Our proposed protocol can detect this because the authenticator always generates new CRPs and does not repeat them. Consequently, if a previous response is sent to the authenticator, it recognizes this as an attack. Hence, the proposed protocol prevents RA.

\paragraph{Device Impersonation (DI) Attacks} 
Since our proposed authentication model does not store CRPs in a database, and the attackers can not access the LPUF-AuthNet models, it is difficult for malicious entities to fabricate a counterfeit PUF using a machine-learning approach. This attribute contributes to the robustness of our method against the DI.

\paragraph{ML Attacks} 
In our proposed scheme, we do not store the CRP dataset, and each time our protocol generates a new CRP. Consequently, the attacker can only store the CRPs used in the previous authentication session and cannot create a model that predicts the next response using a novel challenge. Our model is resilient to ML attacks.

\section{Performance Evaluation}
The performance evaluations of the LPUF-AuthNet architecture are conducted in this section. The CRP dataset used to train the LPUF-AuthNet models is collected from a physical ring-oscillator PUF designed at Khalifa University (KU) \cite{abulibdeh2024security}. The model is trained using the adaptive moment estimation (ADAM) algorithm for efficient convergence. For reproducibility of the results, our source code is made publicly available\footnote{\url{https://github.com/BrahiM-Mefgouda/LPUF-AuthNet}}. 

\subsection{ML-Based Modeling}
To assess the security of our protocol against ML attacks, we evaluate its resistance to two widely used models: support vector machines (SVM) with radial basis function kernel and NN with a 4-layer architecture (64, 32, 16, 8 neurons). These models attempt to predict the LR from incoming LC, simulating an $Att$ trying to mimic a legitimate node. The models collect LC-LR pairs transmitted between the verifier and legitimate nodes and then test the protocol's resilience to sophisticated ML-based attacks.

\begin{figure}[!htbp]
\begin{center}
\includegraphics[width= 1\linewidth]{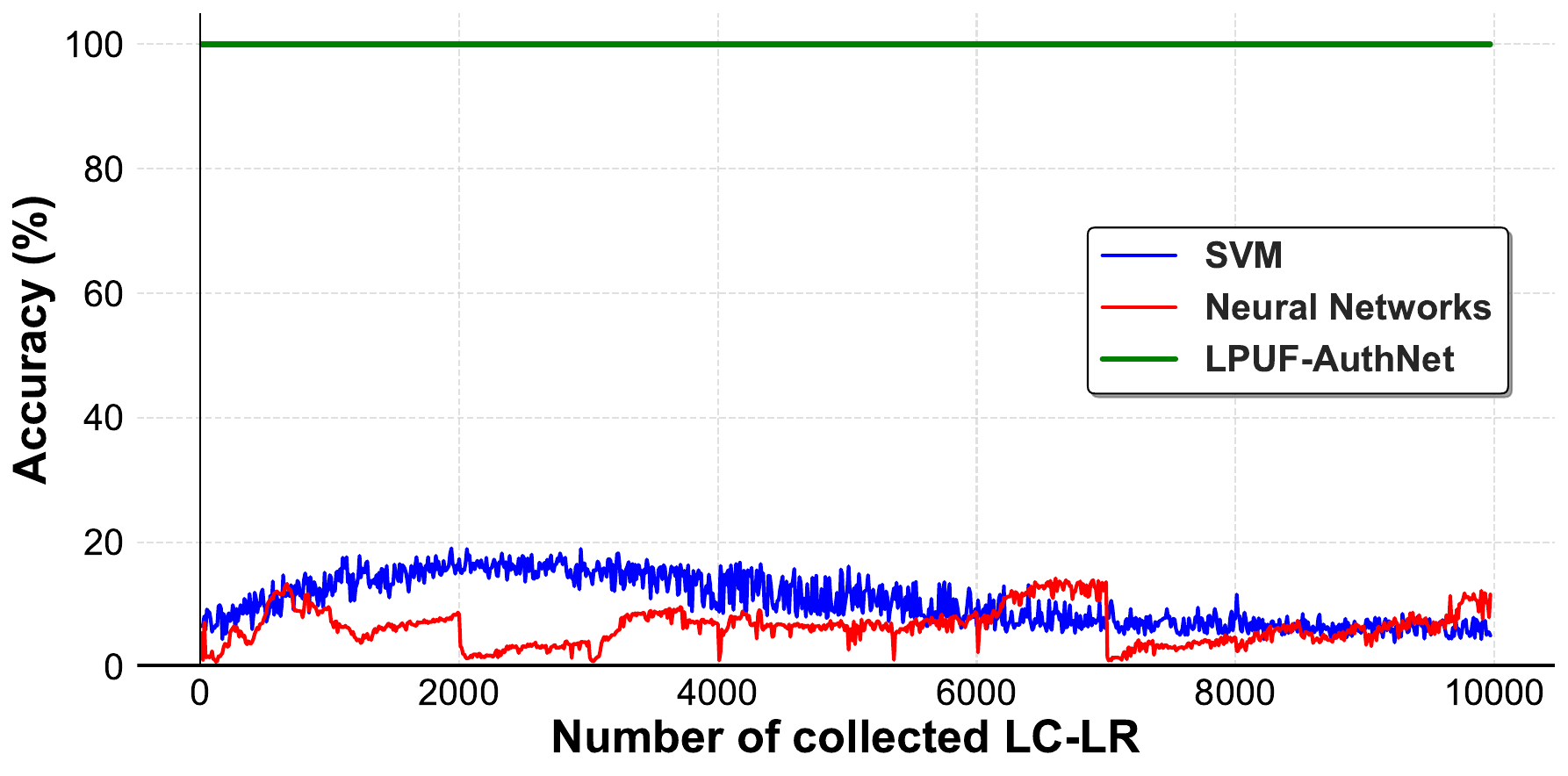}
\caption{Model accuracy comparison: SVM, NN, and LPUF-AuthNet.}
\label{X}       
\end {center}
\end{figure}

{\color{black}
Fig. \ref{X} illustrates the accuracy of predicting LRs after training the models on collected LC-LR pairs. The results demonstrate that both SVM and NN  fail to accurately predict LRs after training, whereas our model consistently produces correct LR predictions. This discrepancy arises from the inability of the SVM and NN models to effectively learn the relationship between LC and LR. Our protocol continually generates novel LC-LR pairs during each authentication session, which prevents the attacker from learning a consistent pattern. In contrast, our proposed protocol fine-tunes all the LC-LR pairs in the dataset, allowing it to accurately predict the LR for the corresponding LC. This low prediction accuracy for SVM and NN underscores our protocol's resistance to ML-based attacks, as higher prediction accuracy would indicate vulnerability. Conversely, our model's superior performance highlights its effectiveness in generating secure, unpredictable responses.

}

\subsection{Latent Challenge Authentication Accuracy}

\begin{figure}[!htb]
\begin{center}
\includegraphics[width= 0.6\linewidth]{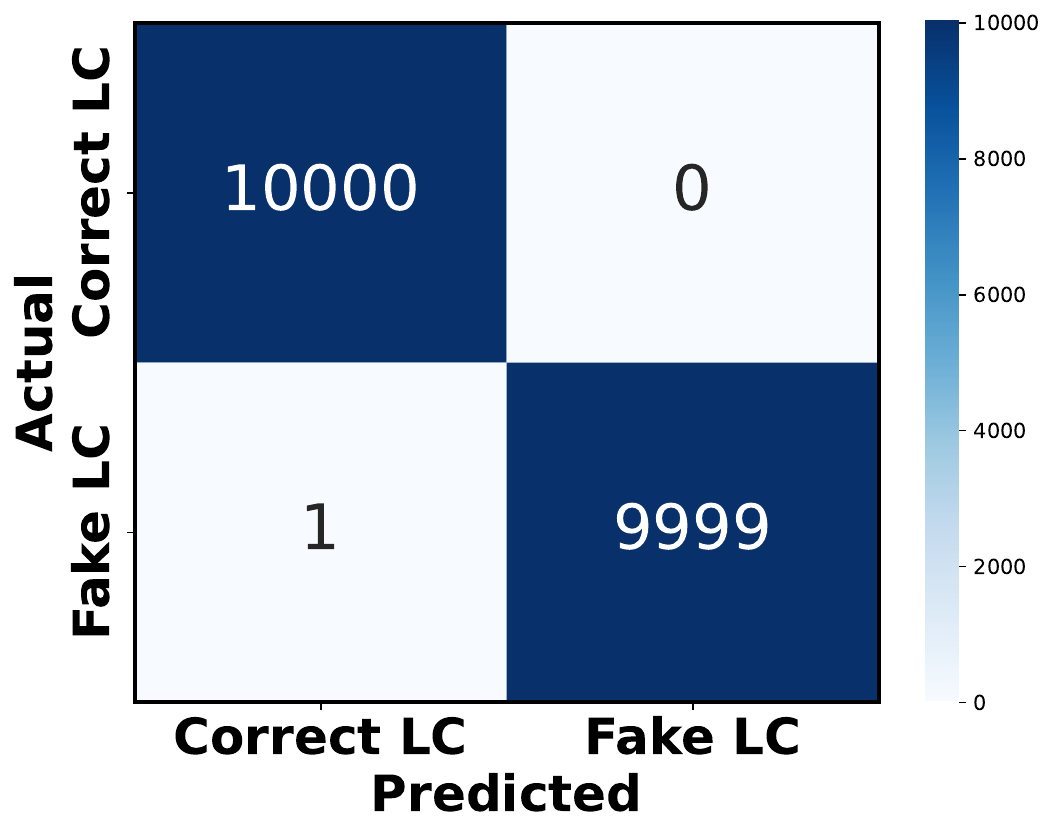}
\caption{Confusion matrix for LC authentication accuracy.}
\label{Matrix}       
\end {center}
\end{figure}

{\color{black}
The robustness of the LPUF-AuthNet system in distinguishing between authentic and fake LCs is evaluated using a comprehensive dataset of $10^4$ LCs. These LCs were derived from the collected dataset and processed through $Decoder_1$. Concurrently, an equal number of random fake LCs were generated to simulate potential adversarial inputs, providing a stringent test of the system's discriminative capabilities.

As illustrated in Fig. \ref{Matrix}, The results demonstrate the exceptional accuracy of the TNN$_2$ component in authentication. Specifically, it achieved a 100\% detection rate for correct LCs and a 99.99\% detection rate for fake LCs. This accuracy in detecting both authentic and counterfeit LCs underscores the effectiveness of the proposed framework in establishing a reliable authentication mechanism for IoT systems
}

\subsection{Data Transmission Overhead}

We evaluate our proposed protocol in terms of number of messages and bits exchanged between the verifier and legitimates nodes compared to existing lightweight protocols, including Chatterjee \cite{chatterjee2018building}, Harishma \cite{harishma2020safe}, Nimmy \cite{nimmy2023novel}, and Zhang \cite{zhang2024building}. Assuming standard bit sizes for various components (identities: 8, random number: 128, challenge: 32, response: 16, hash function: 128, timestamp: 48, zero-knowledge proof of knowledge: 200, nonce: 32). 

\begin{figure}[!htb]
\begin{center}
\includegraphics[width= 1\linewidth]{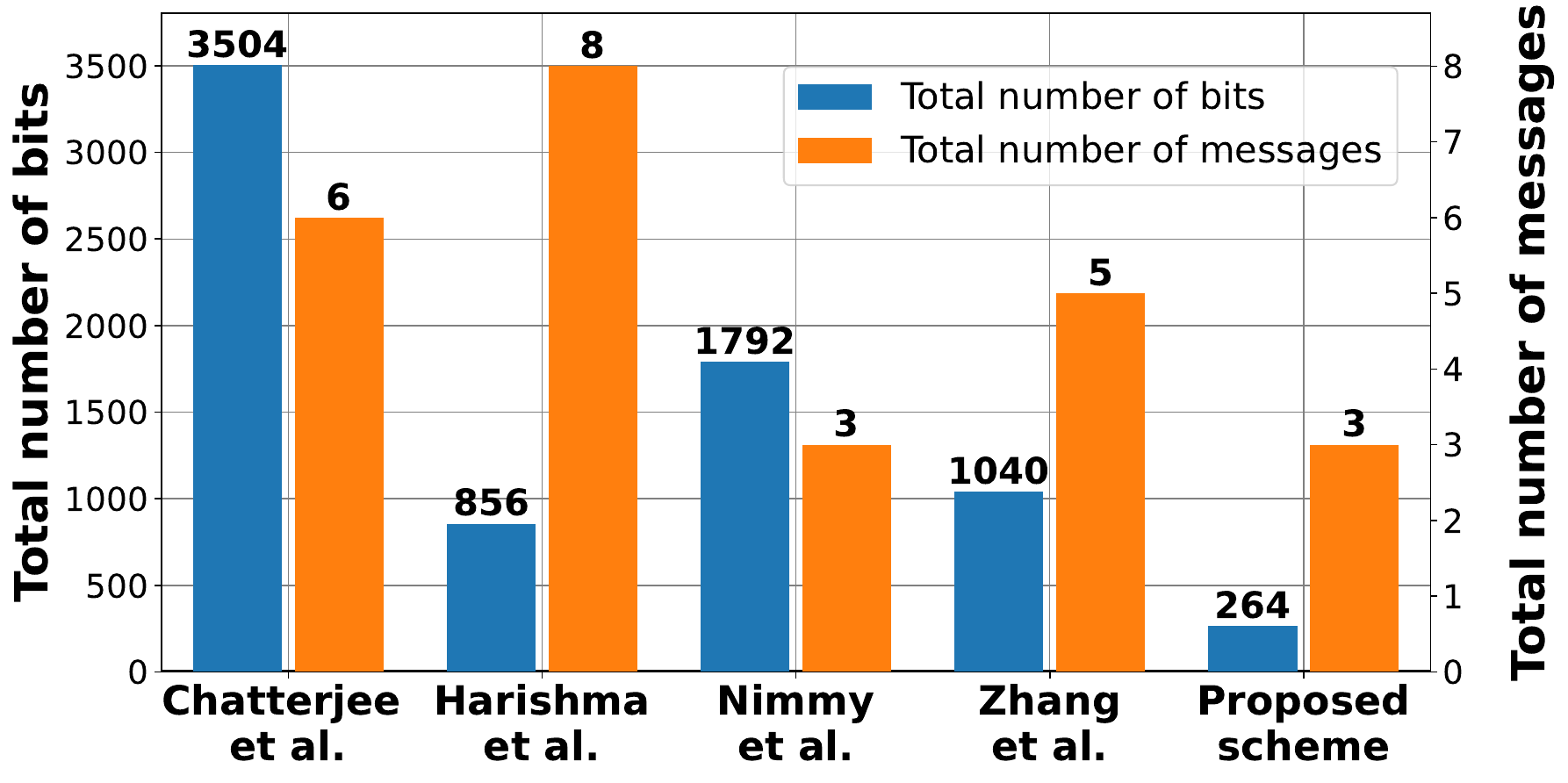}
\caption{Number of messages and bits exchanged during the authentication.}
\label{Numberofmessages}       
\end {center}
\end{figure}

The message exchange counts for Chatterjee, Harishma, Nimmy, and Zhang are 6, 8, 3, and 5, respectively, as illustrated in Fig. \ref{Numberofmessages}, while our protocol uses only 3 messages. In terms of bits, Chatterjee, Harishma, Nimmy, and Zhang require 3504, 856, 1792, and 1040 bits, respectively. In contrast, our protocol uses just 264 bits, including 8 bits for identification, 128 bits for the LC, and 128 bits for the LR. Thus, our protocol is more efficient in terms of message and bit exchanges.

\subsection{Implementation on Raspberry Pi}

We implement the proposed scheme on two Raspberry Pi model B devices (64-bit quad-core Cortex-A72, 4GB RAM). The first device, RasPi$_1$, acts as the verifier, while the second device, RasPi$_2$, acts as the legitimate node, connected via a local wireless network (with TCP/IP). LPUF-AuthNet training is conducted on a Dell laptop with an Intel® Core™ i7-1365U processor and 16GB RAM.

After evaluating the LPUF-AuthNet models with $3\times10^3$ samples, our proposed scheme achieves 99.99\% accuracy with a mean absolute error (MAE) of $2.33 \times 10^{-5}$. and a bit-flip rate of $10^{-3}$\%. These findings validate the robustness and high reliability of our models.

Fig. \ref{A} illustrate the efficacy of the proposed protocol execution. The protocol begins with the authenticator initiating the process while the legitimate node submits an authentication request. The final determination of the node’s legitimacy is then made. The total duration of the authentication process ranges from 1.5 to 1.7 seconds. The size of the dataset stored in the authenticator is approximately 94 KB. However, with our proposed protocol, the authenticator only needs to store DNN $2\times10^3$ KB, and the TNN$_1$ does not exceed $3\times10^3$ KB. Thus, our method saves space compared to methods that store the CRP dataset in the authenticator node.

\begin{tcolorbox}[
    colback=terminalbackground,
    colframe=white,
    leftrule=0.0mm,
    rightrule=0.1mm,
    toprule=0.01mm,
    bottomrule=0.1mm,
    arc=0mm,
    boxsep=1mm,
    left=-0.1mm,
    right=-1mm,
    top=-2.5mm,
    bottom=-2.5mm,
]
\begin{lstlisting}[style=terminal, basicstyle={\fontsize{6}{6}\selectfont\ttfamily\color{white}}]
Hello, this is the authenticator node
Press <g>Enter</g> to continue or <r>exit</r> to quit:
Continuing...
The Authenticator select the Challenge: [[1 1 0 1 1 1 0 1 0 0 0 1 1 1 0 0 0 1 1 1 0 1 1 1 0 1 0 0 0 1 0 0]]
Its corresponding response is: [[0 0 1 0 0 0 0 1 0 0 0 0 1 0 0 1]]
Loading the Encoder1 and Deccoder1 models
1/1 [==============================] - 0s 284ms/step
Encoder1: The LC of the selected challenge is: [0.22, 0.1, 0.9, 0.4]
The authenticator is waiting for a node at 0.0.0.0:8000
The authenticator connects with node: 192.168.137.8:38222 to send the LC
The authenticator received the following LR from the node : [0.23, 0.97, 0.2, 0.002]
The predict response of the received LR is: [[0 0 1 0 0 0 0 1 0 0 0 0 1 0 0 1]]
the Hamming distance between the real response and the predicted response is: 0
The node that sent the response is Legitimate
\end{lstlisting}
\end{tcolorbox}

\begin{tcolorbox}[
    colback=terminalbackground,
    colframe=white,
    leftrule=0.1mm,
    rightrule=0.1mm,
    toprule=0.01mm,
    bottomrule=0.1mm,
    arc=0mm,
    boxsep=1mm,
    left=-0.1mm,
    right=-1mm,
    top=-2.5mm,
    bottom=-2.5mm,
]

\begin{lstlisting}[style=terminal, basicstyle={\fontsize{6}{6}\selectfont\ttfamily\color{white}}]
Hello, this is the legitimate node
The legitimate node sends an authentication request to the authenticator with IP: 192.168.137.253:8000
The node received from the authenticator the following LC: [0.22, 0.1, 0.9, 0.4]
Loading the = model: Encoder2, Decoder2, and Decoder1 
The Hamming distance between the received C' and C" is: 0
The sender is an authenticator
1/1 [==============================] - 0s 279ms/step
Encoder2: The predicted LR is: [0.23, 0.97, 0.2, 0.002]
Listening on 0.0.0.0:8001 to send back the LR to the authenticator
Accepted connection with the authenticator: 192.168.137.253:37968
The node sent the predicted LR to the authenticator
\end{lstlisting}
\end{tcolorbox}
\captionof{figure}{Display Screen of the authenticator  and legitimate nodes during the authentication}
\label{A}
\label{fig:auth_process}
\vspace{2mm}

{\color{black} 
Our proposed protocol maintains its efficiency when expanding the network proving the scalability of the approach. The addition of a legitimate IoT node does not significantly impact the overall system size. This scalability is achieved by only requiring the addition of an IoT device with an instance of TNN$_2$ for each new node. This approach ensures that system growth remains manageable and does not lead to exponential increases in storage or computational requirements.
}

\section{Conclusion}
This paper introduced LPUF-AuthNet, a novel lightweight authentication framework that leverages the unique properties of PUFs while utilizing advanced ML techniques. By integrating DNN within an SL paradigm, LPUF-AuthNet effectively emulates the behaviour of hardware PUFs, reducing communication and overhead and providing robust resistance against various security attacks.
Experimental results demonstrated the efficacy of our protocol. Moreover, a proof-of-concept was developed and demonstrated near-perfect accuracy in detecting fake latent space challenges, validating the practicality and effectiveness of this approach.


\balance
\bibliographystyle{IEEEtran}
\bibliography{bib.bib}
\end{document}